\documentstyle[12pt]{article}
\input epsf
\begin{document}
\title{Polarised parton densities from the fits to the deep inelastic spin
asymmetries  on nucleons}
\author{ Jan
Bartelski\\ Institute of Theoretical Physics, Warsaw University,\\
Ho$\dot{z}$a 69, 00-681 Warsaw, Poland. \\ \\ \and Stanis\l aw
Tatur
\\ Nicolaus Copernicus Astronomical Center,\\ Polish Academy of
Sciences,\\ Bartycka 18, 00-716 Warsaw, Poland. \\ }
\date{}
\maketitle
\vspace{1cm}
\begin{abstract}
\noindent We have made next to leading order QCD fit to the deep
inelastic spin asymmetries on nucleons and we have determined
polarised quark and gluon densities. The functional form for such
distributions was inspired by the unpolarised ones given by MRST
(Martin, Roberts, Stirling and Thorne) fit. In addition to usually
used data points (averaged over $Q^2$) we have also considered the
sample containing points with the same $x$ and different $Q^2$.
Our fits to both groups of data give very similar results. For the
integrated quantities we get rather small gluon polarisation. For
the non averaged data the best fit is obtained with vanishing
gluon contribution at $ Q^{2}=1\, {\rm GeV^{2}}$. For comparison
models with alternative assumptions about quark sea and in
particular strange sea behaviour  are discussed.
\end{abstract}

\newpage
One has quite a lot of data on deep inelastic spin asymmetries for
different nucleon targets. The data (recent and old) come from experiments
made at SLAC [1-9],  CERN [10-15] and DESY \cite{hermes,hernew}.
The newest data on proton \cite{e143new,SMCnew,hernew} and deuteron
targets \cite{e143new,e155,SMCnew} have smaller statistical errors and
hence dominate in $\chi^2$ fits.

Since the analysis of the EMC group results \cite{emc} an enormous
interest were started in studying polarised structure functions.
It was suggested \cite{dg} that polarised gluons may be
responsible for the little spin carried by quarks. Determination
of polarised gluon distribution is particularly interesting in
this context. After calculation of two loop polarised splitting
functions \cite{ew} several next to leading (NLO) order QCD fits
were performed [20-24] and polarised parton distributions (i.e.
for quarks, antiquarks and gluons) were determined. Many groups
obtained rather highly polarised gluon contribution (however this
number is determined with a big error). The aim of this paper is
to extend our next to leading order QCD analysis given in
\cite{BT} by taking into account, in addition to all previously
considered data, also proton data from Hermes (DESY) and deuteron
data from E155 experiment (SLAC). We will also use more recent
MRST fit for partons \cite{MRSTnew}. In \cite{BT} we got
negligible gluon contribution at $ Q^{2}=1\, {\rm GeV^{2}}$. We
will see that our result about gluon polarisation does not change
very much. The method of choosing basic functions for fitting,
used in this paper, differs from the other groups. As in \cite{BT}
we will divide the data into two groups. Let us remind that many
experimental groups published data \cite{e143new,SMCnew} sets for
the near values of $x$ and different $Q^2$ in addition to the
"averaged" data where one averages over $Q^2$ (the errors are
smaller and $Q^2$ dependence is smeared out). In principle when we
take into account $Q^2$ evolution of polarised and unpolarised
functions (in NLO analysis) the first group of data points, i.e.
non averaged, should be considered. In most of the fits to
experimental data only second group of data, namely with averaged
$Q^2$ dependence, was used. We will make fits to the both sets of
data (the first group contains 417 points and the second 159
points). The results for both groups of data are very similar (the
same conclusion was already drawn in \cite{BT}). We will also
compare those results with the fits without $Q^2$ evolution taken
into account (in other words assuming that asymmetries do not
depend on $Q^2$, as in our previous fits \cite{bt1}). One should
add that many experimental groups have not succeeded in finding
$Q^2$ dependence for approximately the same value of $x$ and
different $Q^2$ \cite{e143qd,SMCqp,SMCqd}. In our analysis we
limit ourselves, as one usually does, to the data with $Q^2 \geq 1
\mbox{GeV}^2$. As was already mentioned in our earlier papers
\cite{bt1} making a fit to spin asymmetries and not directly to
$g_1 (x,Q^2)$ enables to avoid the problem with higher twist
contributions which are probably less important in such case ( see
for example \cite{man}) . Experiments on unpolarised targets
provide information on the unpolarised quark densities $q(x,Q^2)$
and $G(x,Q^2)$ inside the nucleon. These densities can be
expressed in term of $q^{\pm}(x,Q^2)$ and $G^{\pm}(x,Q^2)$, i.e.
densities of quarks and gluons with helicity along or opposite to
the helicity of the parent nucleon. The unpolarised quark
densities are given by the sum of $q^{+}$, $q^{-}$ and $G^{+}$  ,
$G^{-}$, namely:
\begin{equation}
q = q^{+}+q^{-},\hspace*{1.5cm} G = G^{+}+G^{-}.
\end{equation}

On the other hand the polarised DIS experiments give information about
polarised parton densities, i.e. the difference of $q^{+}$, $q^{-}$
and $G^{+}$, $G^{-}$:
\begin{equation}
\Delta q = q^{+}-q^{-},\hspace*{1.5cm} \Delta G = G^{+}-G^{-}.
\end{equation}

We will try to determine $q^{\pm} (x,Q^2)$ and $G^{\pm} (x,Q^2)$,
in other words, we will try to connect unpolarised and polarised
data. In principle the asymptotic $x$ behaviour of $q^{+}$ and
$q^{-}$ will be taken from the unpolarised data (up to the
modifications when some leading order terms vanish). We will use
the fit given by MRST \cite{MRSTnew}. In comparison with their
previous fit (called R2) \cite{mrsn} there is different behaviour
at small $x$ for quark and gluon distributions. It is of course
very restrictive assumption that $\Delta q$ and $\Delta G$ have
the same behaviour as $q$ and $G$. On the other hand the small $x$
behaviour of unpolarised structure functions is determined from
the $x$ values of Hera much smaller than in the polarised case. In
our further analysis we will consider the integrals over the
region measured in the experiments with polarised particles with
the hope that in this case the behaviour of $q^{\pm}$ and
$G^{\pm}$ could be more plausible. The values of integrals in the
whole region ($0 \leq x \leq 1$), involving asymptotic behaviour
taken from the unpolarised structure functions, may be not as
reliable as in the measured region. As an alternative we also use
Regge type behaviour.

It is known \cite{grv} that the behaviour of the quark and gluon
distributions in small $x$ region is extremely important in
extrapolation of integrals over whole $ 0 \leq x \leq 1$ range. It
could happen that in $\Delta q = q^{+} - q^{-}$ (when we assume
that $q^{+}$, $q^{-}$ and $q = q^{+} + q^{-}$ have similar $x$
dependence) most singular $x$ terms cancel (that is especially
important in case of  valence quarks and sea contributions where
the $x$ behaviour is relatively singular). We will see later how
such description infers the fits and calculated parameters. For
the less singular  distributions for $\Delta u_v$, $\Delta d_v$
and $\Delta M$ (total sea polarisation) there is no strong
dependence of calculated quantities on the extrapolation in an
unmeasured region but the fits have higher $\chi^2$. One of the
main tasks of considering NLO evolution in $Q^2$ is the
determination of the gluon contribution $\Delta G$. In
$\overline{MS}$ scheme $\Delta G(x,Q^2)$ comes in through the
higher order
 corrections. In our fits we obtain $\Delta G$ relatively small contrary to
some expectations. When we use the non-averaged sample of data the
actually the  best fit is with vanishing $\Delta G$ contribution.

Let us start with the formulas for unpolarised quark parton
distributions gotten (at $ Q^{2}=1\, {\rm GeV^{2}}$) from the one of
recent fits performed by Martin, Roberts, Stirling and Thorne
\cite{MRSTnew}. For the valence
quarks one get (in this fit one uses
$\Lambda^{n_{f}=4}_{\overline{MS}}=0.3$
$\mbox{GeV}$ and $\alpha_s(M^2_Z)=0.120$):
\begin{eqnarray}
u_{v}(x)&=&0.6051
x^{-0.5911}(1-x)^{3.395}(1+2.078\sqrt{x}+14.56x), \nonumber \\
d_{v}(x)&=&0.0581 x^{-0.7118}(1-x)^{3.874}(1+34.69\sqrt{x}+28.96x)
,
\end{eqnarray}

\noindent and for the antiquarks from the sea (the same distribution is
for sea quarks):
\begin{eqnarray}
2\bar{u} (x)&=&0.4M(x)-\delta (x), \nonumber \\
2\bar{d} (x)&=&0.4M(x)+\delta (x),  \\
2\bar{s} (x)&=&0.2M(x). \nonumber
\end{eqnarray}

\noindent In eq.(4) the singlet contribution $M=2[\bar{u}+\bar{d}
+\bar{s}$] is:
\begin{equation}
M(x)=0.2004 x^{-1.2712}(1-x)^{7.808}(1+2.283\sqrt{x}+20.69x),
\end{equation}

\noindent whereas the isovector part ($\delta=\bar{d}-\bar{u}$) reads:
\begin{equation}
\delta (x)=1.290 x^{0.183}(1-x)^{9.808}(1+9.987x-33.34x^{2}).
\end{equation}

\noindent For the unpolarised gluon distribution one gets:
\begin{equation}
G(x)=64.57 x^{-0.0829}(1-x)^{6.587}(1-3.168\sqrt{x}+3.251x).
\end{equation}

We assume, in an analogy to the unpolarised case, that the
polarised quark distributions are of the form:
$x^{\alpha}(1-x)^{\beta}P( \sqrt{x})$, where $P(\sqrt{x})$
is a polynomial in $\sqrt{x}$ and the asymptotic
behaviour for $x \rightarrow 0$ and $x \rightarrow 1$ (i.e. the
values of $\alpha$ and $\beta$) are the same (except for $\Delta
M$, see a discussion below) as in the unpolarised case. Our idea is to
split the numerical constants (coefficients of $P$ polynomial)
in eqs.(3, 5, 6 and 7) in two parts in such a manner that the
distributions $q^{\pm}(x,Q^2)$ and
$G^{\pm}(x,Q^2)$ remain positive. At the end of the paper we will
discuss the consequences of relaxing the positivity conditions. Our
expressions for $\Delta q(x) = q^{+}(x)-q^{-}(x)$ ($q(x) =
q^{+}(x)+q^{-}(x)$) are:
\begin{eqnarray}
\Delta u_{v}(x)&=&x^{-0.5911}(1-x)^{3.395}(a_{1}+a_{2}\sqrt
{x}+a_{4}x), \nonumber \\ \Delta
d_{v}(x)&=&x^{-0.7118}(1-x)^{3.874}(b_{1}+b_{2}\sqrt{x}+b_{3}x),
\nonumber \\ \Delta
M(x)&=&x^{-0.7712}(1-x)^{7.808}(c_{1}+c_{2}\sqrt{x}), \\ \Delta
\delta (x)&=&x^{0.183}(1-x)^{9.808}c_{3}(1+9.987x-33.34x^{2}),
\nonumber
\\ \Delta G (x)&=&x^{-0.0829}(1-x)^{6.587}(d_1+d_2 \sqrt{x}+d_3 x).
\nonumber
\end{eqnarray}

It is very important what assumptions one makes about the sea
contribution. From the MRST fit for unpolarised structure
functions the natural assumption would be: $\Delta \bar{s}=\Delta
\bar{d}/2 = \Delta \bar{u}/2$. If we add the condition that SU(3)
combination: $a_8=\Delta u+\Delta d-2 \Delta s$ should be equal
to the value determined from the semileptonic hyperon decays,
$\Delta s$ is pushed into negative values and so is nonstrange sea.
Instead of connecting
$\Delta s$ in some way to non-strange sea value we introduce
additional free parameters for the strange sea contribution namely
\begin{equation}
\Delta M_s= x^{-0.7712}(1-x)^{7.808} (c_{1s}+c_{2s}\sqrt{x}) .
\end{equation}
For the strange quarks we have additional independent
parameters. Hence, in our fits we will start with fourteen parameters.
Comparing the expression (5) with (8) and (9) we see that in
$\Delta M$ (and $\Delta M_s$) there is no term behaving like
$x^{-1.2712}$ at small
$x$ (hence, we assume that $\Delta M $  and hence all sea distributions
have finite integral) which means that in $\Delta M$  coefficient in
front of this term have to be splitted into equal parts in $\Delta
M^+$ and $\Delta M^-$ (the most singular term
in the sea contribution drops out). Hence, in the fitting procedure
we are using functions that are suggested by the fit to unpolarised
data. Maybe not all of them are important in the fit and it could
happen that some of the coefficients in eqs.(8,9) taken as free
parameters in the fit are small or in some sense superfluous.
Putting  them to zero or eliminating them increase $\chi^{2}$ only
a little but makes $\chi^{2}/N_{DF}$ smaller. We will see that
that is the case with some parameters introduced in eqs. (7,8).
On the other hand we have still relatively strong
singular behaviour of $\Delta u_v$ and $\Delta M$  for small $x$
values. For comparison we will also consider later the model in which
most singular terms are put equal to zero namely
$a_1=c_1=c_{1s}=0$, which means that plus and minus components have the
same coefficients for this power of $x$. In the less singular models
the dependence of calculated
parameters in the unobserved region (below $x \leq 0.003$) is
weak. In the earlier papers we considered the extrapolation of
various calculated integrals below $x=0.003$ up to 0 assuming
Regge type of behaviour for small $x$ values. As will be discussed
later the less singular models give significantly higher $\chi^2$.

In order to get the unknown parameters in the expressions for polarised
quark and gluon distributions (eqs.(8,9)) we calculate the spin asymmetries
starting from initial $Q^2$ = 1 $\mbox{GeV}^2$ for measured values
of $Q^2$ and make a fit to the experimental data on spin asymmetries for
proton, neutron and deuteron targets. The asymmetry $A_1(x,Q^2)$ can
be expressed via the polarised structure function $g_1(x,Q^2)$ as
\begin{equation}
A_1(x,Q^2)\cong \frac{(1+\gamma^{2}) g_{1}(x,Q^2)}{ F_1(x,Q^2)}=
\frac{ g_{1}(x,Q^2)}{ F_2(x,Q^2)}[2x(1+ R(x,Q^2))]
\end{equation}
\noindent where  $R = [F_2(1+\gamma^{2})-2xF_1]/ 2xF_1$ whereas
$F_1$ and $F_2$ are the unpolarised structure functions and
$\gamma =2Mx/Q$. We will take the new determined value of $R$
from the \cite{whitn}. The factor $(1+\gamma^{2})$ plays non negligible
role for $x$ and $Q^{2}$ values measured in SLAC experiments.
 Polarised structure function $g_1(x,Q^2)$ in the next to leading order
QCD is related to the polarised quark and gluon distributions
$\Delta q(x,Q^2)$, $\Delta G(x,Q^2)$, in the following way:
\begin{eqnarray}
g_{1}(x,Q^2)&= \frac{1}{2} \sum_{q}e_{q}^2\{\Delta
q(x,Q^2)+\frac{\alpha_s}{2 \pi}[\delta c_q *\Delta q(x,Q^2)
\nonumber \\ &+\frac{1}{3}\delta c_g* \Delta G(x,Q^2)]\}
\end{eqnarray}
\noindent
with the convolution * defined by:
\begin{equation}
(C*q)(x,Q^2) = \int_{x}^{1}\frac{dz}{z} C(\frac{x}{z})q(z,Q^2)
\end{equation}

The explicit form of the appropriate spin dependent Wilson coefficient
$\delta c_q$ and $\delta c_g$ in
the $\overline{MS}$ scheme can be found for example in ref. \cite{ew}. The
NLO expressions for the unpolarised (spin averaged) structure function
is similar to the one in eq.(11) with $\Delta q(x,Q^2)
\rightarrow q(x,Q^2)$ and the unpolarised Wilson coefficients are given in
\cite{ewol1,ewol2}.

The $Q^2$ evolution of the parton densities is governed by the DGLAP
equations \cite{glap}. For calculating the NLO evolution of the spin
dependent parton distributions $\Delta q(x,Q^2)$, $\Delta G(x,Q^2)$ and
spin averaged $q(x,Q^2)$, $G(x,Q^2)$ ones we will follow the method
described in \cite{grv,ewol2}. We will
calculate Mellin n-th moment of parton distributions $\Delta q(x,Q^2)$ and
$\Delta G(x,Q^2)$ according to:
\begin{equation}
\Delta q^{(n)}(Q^2)=\int_0^1 dx x^{n-1} \Delta q(x,Q^2)
\end{equation}
and then use NLO solutions in Mellin n-moment space in order to calculate
evolution in $Q^2$ for non-singlet and singlet parts.

In calculating evolution of $\Delta \Sigma^{(n)}(Q^2)$ and $\Delta
G^{(n)}(Q^2)$ with $Q^2$ we have
mixing governed by the anomalous dimension 2x2 matrix \cite{ewol2}. Having
evolved moments one can insert them
into the n-th moment of eq.(11).
\begin{eqnarray}
g^{(n)}_{1}(Q^2)&=\frac{1}{2} \sum_{q}e_{q}^2 \{\Delta
q^{(n)}(Q^2)+\frac{\alpha_s}{2 \pi}[\delta c_{q}^{(n)} \cdot\Delta
q^{(n)}(Q^2)  \nonumber \\ &+\frac{1}{3}\delta c_{g}^{(n)}\cdot
\Delta G^{(n)}(Q^2)]\}
\end{eqnarray}
\noindent and then numerically Mellin invert the whole expression.
In this way we get $g_1(x,Q^2)$. The same procedure is applied
for the unpolarised
structure functions. Having calculated the asymmetries according
to equation (10) for the measured in experiments value of $Q^2$ we
can make a fit to a measured asymmetries on proton, neutron and
deuteron targets. We will take into account all together 417 points
(193 for proton, 171 for deuteron and 53 for neutron. We will use
also the "experimental" value  of $a_8 = 0.58 \pm 0.1$ with
enhanced (to 3$\sigma$) error.

The fit with all fourteen parameters from eqs.(8,9) gives
$\chi^{2}=340.4$. It seems that some of the parameters of the most
singular terms are superfluous and we can eliminate them. We will
put $d_{1}=d_{2}=0$ (such assumption gives that $\delta G/G \sim
x$ for small $x$), $b_{1}=0$ (the most singular term in $\Delta
d_{v}$) and assume $c_{1s}=c_{1}$ (i.e. the most singular terms
for strange and nonstrange sea contributions are equal). Fixing
these four parameters in the fit practically does not change
$\chi^{2}$ but improves $\chi^{2}/N_{DF}$. The resulting
$\chi^{2}$ per degree of freedom is better than in the previous
fit and one gets $\chi^{2}/N_{DF}$ =$\frac{341.1}{418-10}$ =0.84.
In this case we get the following values of parameters from the
fit to all existing (above mentioned) data for $Q^2 \geq 1
\mbox{GeV}^2$ for spin asymmetries:
\begin{equation}
\begin{array}{lll}
a_{1}=\hspace*{0.35cm} 0.61 \pm 0.00 ,&a_{2}=-6.1 \pm 0.19
,&a_{4}=\hspace*{0.13cm} 15.7 \pm 0.42,\\ b_{2}\hspace*{0.06cm}=
-1.56 \pm 0.20, &b_{3}\hspace*{0.055cm}=-0.43 \pm 0.49,&\\
c_{1}=-0.40 \pm 0.03,&c_{2}= 4.15 \pm 0.00,\\
c_{1s}\hspace*{0.055cm}=\hspace*{0.00cm}
c_{1},&c_{2s}\hspace*{0.055cm}= -0.28 \pm 0.83,\\
c_{3}\hspace*{0.06cm}=-1.29 \pm 2.53 ,\\ d_3=\hspace{0.13cm} 2.01
\pm 11.2 .&&.
\end{array}
\end{equation}
Actually the parameter  $d_{3}$ could be put equal to zero without
increasing $\chi^2/N_{DF}$. We get in this case the smallest
$\chi^{2}/N_{DF}$ =$\frac{341.1}{418-8}$ =0.83. That means that
$d_3$ is not well determined in the fit and the best
$\chi^2/N_{DF}$ is without gluonic contribution.

The obtained quark and gluon distributions lead for $Q^2$ =1
$GeV^2$ to the following integrated (over $x$) quantities: $\Delta
u = 0.80 \pm 0.02, \Delta d= -0.65 \pm 0.03,
 \Delta s= -0.21 \pm 0.05, \Delta u_v = 0.67 \pm 0.02,
  \Delta d_v = -0.59 \pm 0.02, 2\Delta \bar{u} = 0.14
  \pm 0.03, 2\Delta \bar{d} = -0.07 \pm 0.03.$

These numbers yield the following predictions: $a_0=\Delta
u+\Delta d+\Delta s = - 0.06 \pm 0.07, a_3 =\Delta u -\Delta d =
1.45 \pm 0.02, \Delta G=0.04 \pm 0.19, \Gamma_1^p = 0.111 \pm
0.006, \Gamma_1^n = -0.096 \pm 0.006, \Gamma_1^d = 0.007 \pm
0.005.$

We have positively polarised sea for up and negatively for down
quarks and very strongly negatively polarised sea for strange
quarks. Because of the big negative value of $\Delta s$ the
quantity $a_0$ is negative. The gluon polarisation is small. The
value of $a_3$ was not assumed as an input in the fit (as is the
case in nearly all fits \cite{inni}) and comes out slightly higher
than the experimental value. $\Delta \delta$, comes out relatively
big from the fit (coefficient in front of $\Delta \delta$ is equal
to that in $\delta$).

The asymptotic behaviour at small $x$ of our polarised quark
distributions is determined by the unpolarised ones and these do
not have the expected theoretically Regge type behaviour or pQCD
which is also used by some groups, to extrapolate results to small
values of $x$. Some of the quantities in our fit change rapidly
for $x \leq 0.003$.

Hence, we will present quantities integrated over the region from
$x$=0.003 to $x$=1 (it is practically integration over the region
which is covered by the experimental data, except of non
controversial extrapolation for highest $x$). The corresponding
quantities are $\Delta u = 0.85$ ($\Delta u_v = 0.56$, $2\Delta
\bar{u} = 0.29$), $\Delta d = -0.48$ ($\Delta d_v = -0.57$,
$2\Delta \bar{d} = 0.09$), $\Delta s = - 0.12$, $a_0= 0.25$,
$\Delta G = 0.04$,  $\Gamma_1^p = 0.123$, $\Gamma_1^n = -0.056$,
$\Gamma_1^d = 0.036$, $a_{3}=1.32$. In this region the obtained
values of sea contributions are relatively high and those of
valence quarks relatively small. Gluon contribution practically
vanishes. There is relatively strong dependence of different
quantities in the unmeasured region ($0 \leq x \leq 0.003$). May
be the unpolarised MRST parton distributions ( with the above
mentioned modifications) do not describe quite correctly the small
$x$ behaviour of polarised parton distributions. On the other hand
the fit to the data is very good. So, the values of integrated
quantities in the measured region we consider as more reliable
then in the whole region. With the value of $\Delta s=-0.12$ in
the measured region of $x$ we have $a_0=0.25$ and with $\Delta
s=-0.21$ in the whole region of $x$ $a_0$ becomes negative
(-0.06).

When we use the quantities calculated in the measured region and
extend them to the full $x$ region using asymptotic Regge
behaviour for small $x$ we get $\Delta u = 0.86$ ($\Delta u_v =
0.59$, $2\Delta \bar{u} = 0.27$), $\Delta d = -0.51$ ($\Delta d_v
= -0.58$, $2\Delta \bar{d} = 0.07$), $\Delta s = - 0.14$, $a_0 =
0.21$, $\Delta G = 0.04$, $a_{3}=1.37$. We have used $x^{-\alpha}
$ behaviour for small $x$ (with $-0.25 \leq \alpha \leq 0.25$) and
the quantities do not depend strongly on a specific value of
$\alpha$. For the given above values $\alpha=0$ was used.

\begin{figure}[hb]
\noindent \hspace{-0.5cm} \epsfxsize =410pt\epsfbox{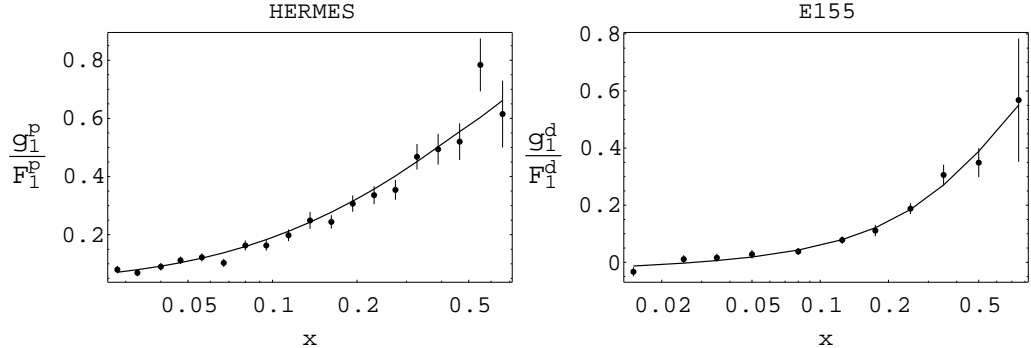}
\caption{\em{The comparison of our predictions for
$g_1^N(x,Q^2)/F_1^N(x,Q^2)$ from basic fit with the recent
averaged data for proton and deuteron targets.}}
\end{figure}

Now, we shall calculate $\Gamma^p$, $\Gamma^n$ and $\Gamma^d$ in
the measured region for $Q^2$ = 5 $\mbox{GeV}^2$ and compare them
with the quantities given by the experimental groups. We get in
the region between $x=0.003$ and $x=0.8$ (covered by the data)
$\Gamma_1^p = 0.132 \pm 0.006$, $\Gamma_1^n = -0.051 \pm 0.007$
and $\Gamma_1^d = 0.037 \pm 0.006$. The experimental group SMC
present \cite{smcth} the following values in such region (for
$Q^2$ = 5 $\mbox{GeV}^2$):
\begin{eqnarray}
\Gamma_1^p &=& \hspace*{0.33cm} 0.130 \pm 0.007, \nonumber \\
\Gamma_1^n &=& -0.054\pm 0.009,\\
\Gamma_1^d &=& \hspace*{0.33cm} 0.036 \pm 0.005. \nonumber
\end{eqnarray}

\noindent One can see that our results are in good agreement with
experimental values. For comparison we have also made fits using
formulas of the simple parton model (as in our papers before
\cite{bt1}) neglecting evolution of parton densities with $Q^2$.
More detailed result of these fits (integrated densities and so
on) will be given later.

In fig.1 as an example we present our fit to the non averaged data
in comparison with measured (averaged over $Q^2$) $g_{1}/F_{1}$
for new proton (Hermes) and deuteron (E155) data. The curves are
obtained by joining the calculated values of asymmetries
corresponding to actual values of $x$ and $Q^2$ for measured data
points. The curves are not fitted but the difference in fitted
asymmetries for averaged and non-averaged data are very small. For
asymmetries the curves with $Q^2$ evolution taken into account and
evolution completely neglected do not differ very much so we do
not present them.

\begin{figure}[hpb]
\noindent \hspace{-0.5cm} \epsfxsize =415pt \epsfbox{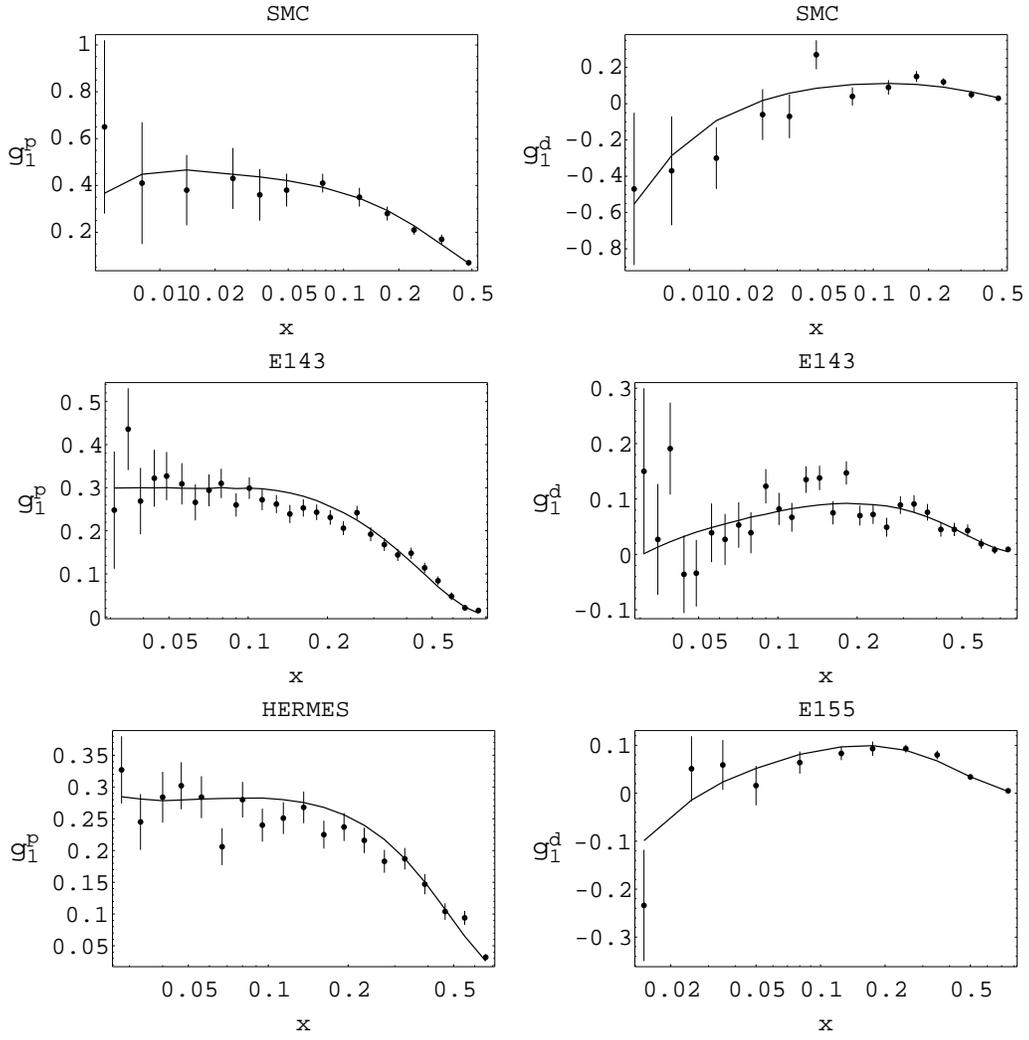}
\caption{\em{The comparison of our predictions for $g_1^N(x,Q^2)$
with the measured structure functions in different experiments and
on different nucleon targets.}}
\end{figure}

\begin{figure}[hpb]
\noindent \hspace{-0.5cm} \epsfxsize =410pt \epsfbox{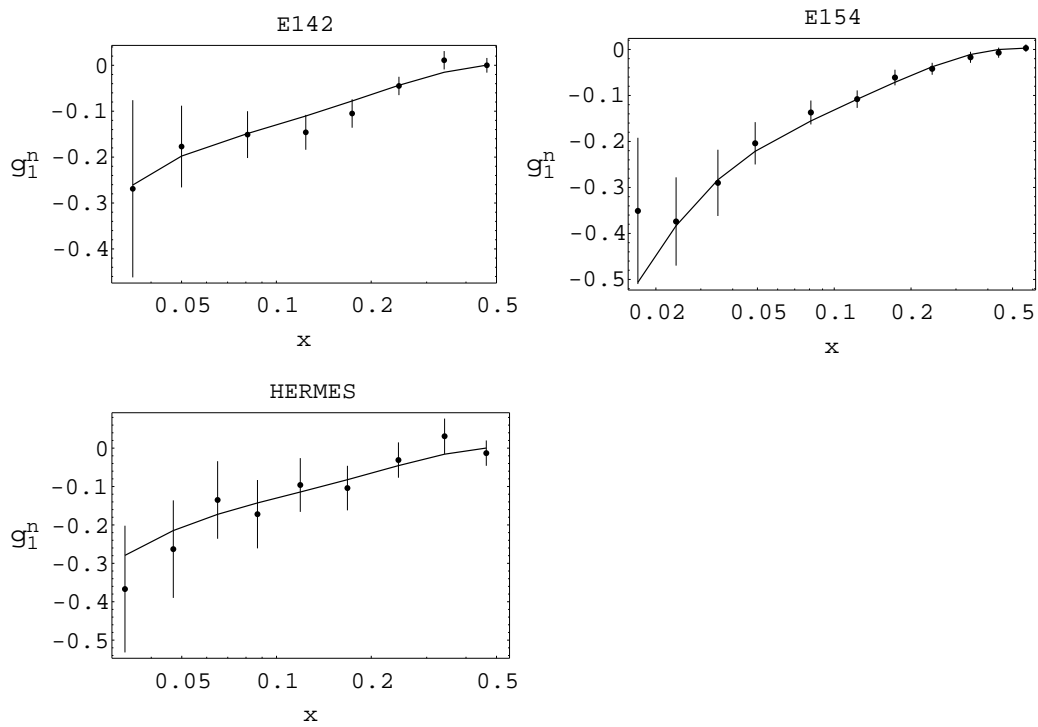}
\caption{\em{The comparison of our predictions for $g_1^n(x,Q^2)$
calculated from basic fit with the measured structure functions in
E142, E155 and Hermes experiments.}}
\end{figure}

\begin{figure}[hpb]
\noindent
\hspace{-0.5cm}
 \epsfxsize =400pt \epsfbox{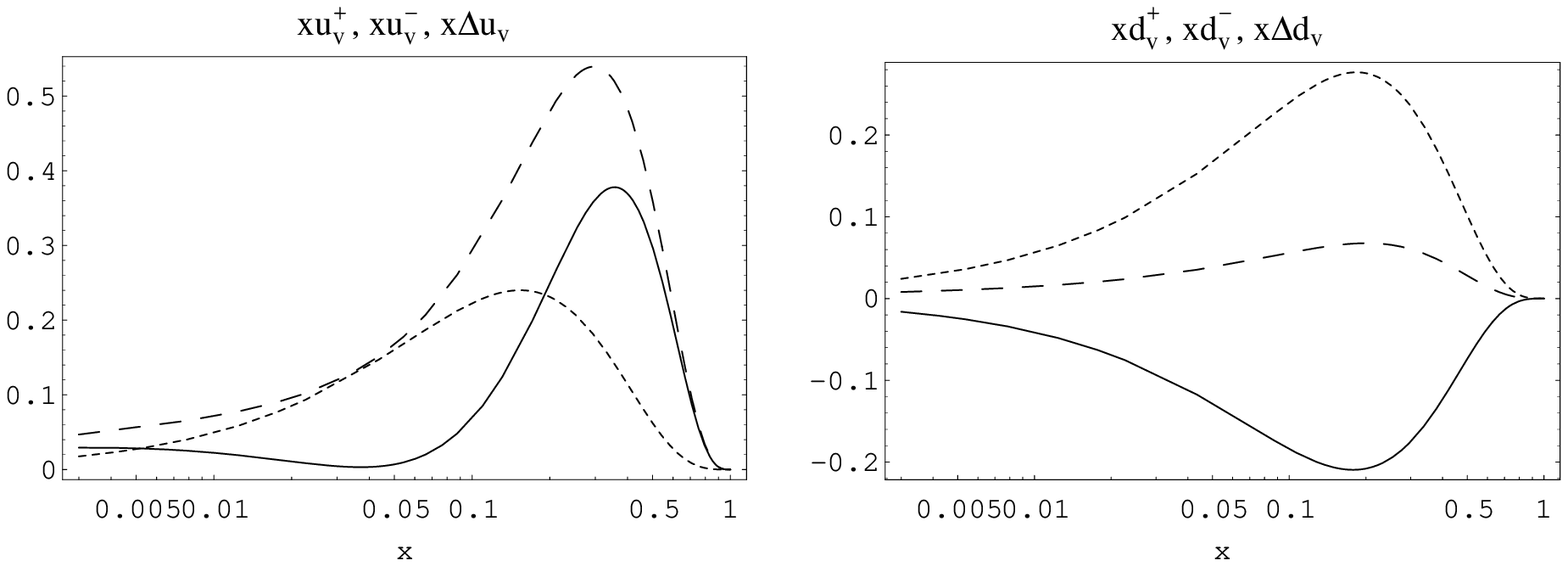}

\noindent \hspace{-0.5cm} \epsfxsize =400pt \epsfbox{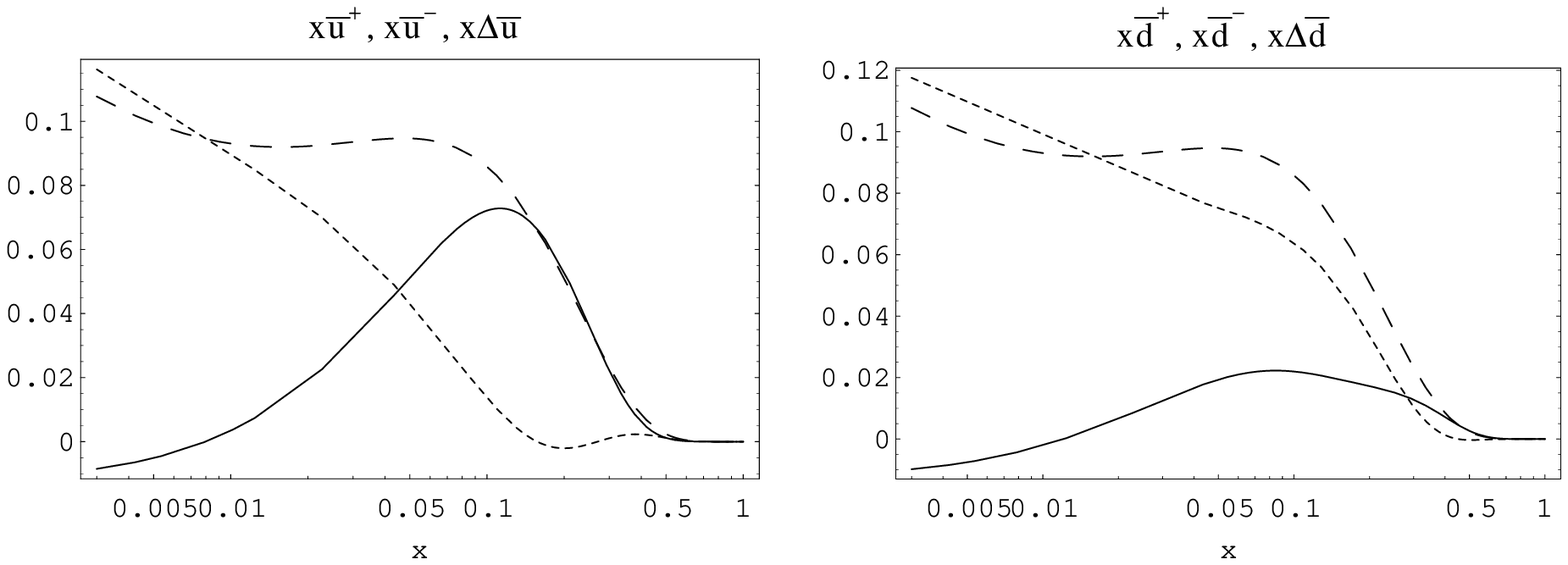}

\noindent \hspace{-0.5cm} \epsfxsize =400pt \epsfbox{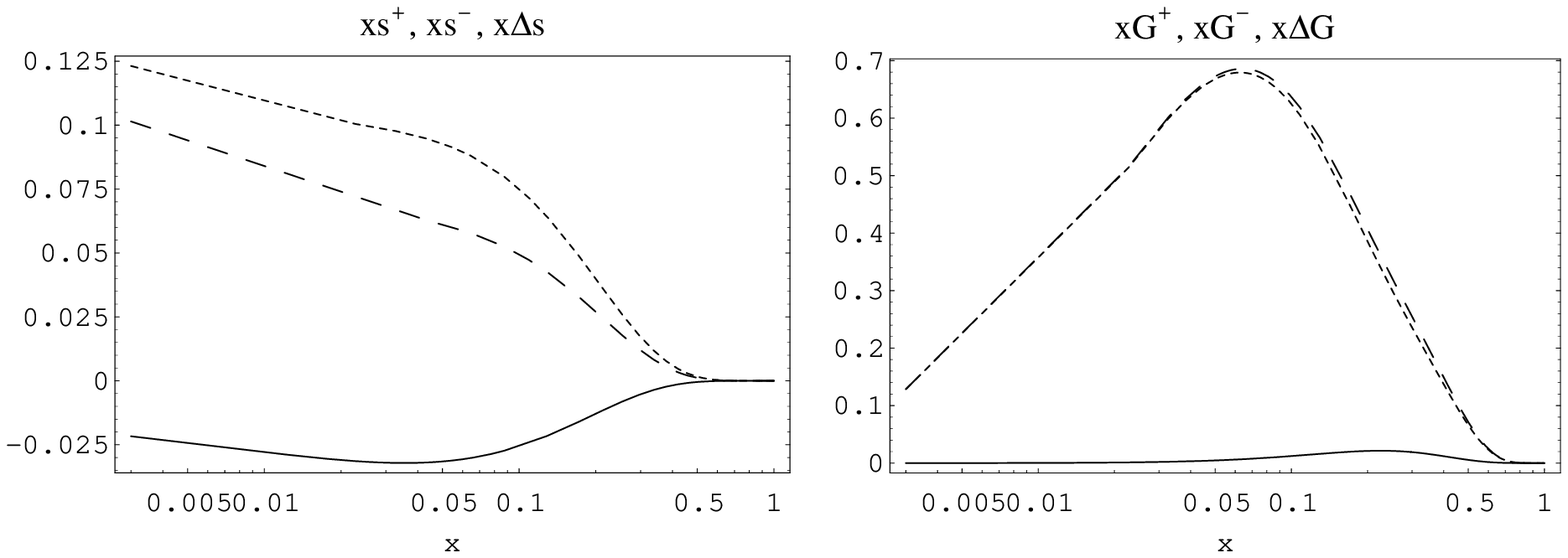}
\caption{\em{Our predictions for spin densities for quark and
gluons at $ Q^{2}=1\, {\rm GeV^{2}}$. We present distributions for
partons polarized along ($xq^+(x),xG^+(x)$) and opposite
($xq^-(x),xG^-(x)$)  to the helicity  of parent proton. The
polarized densities  (i.e. the differences of above mentioned
quantities) are also shown.}}
\end{figure}

In figs. 2 and 3 we show the comparison of our predictions for
$g_{1}$ from the basic fit with the measured averaged values for
proton, deuteron and neutron data. The values of $g_{1}$ were
calculated for the values of $x$ and $Q^2$ measured for averaged
data points in different experiments and then joined together. The
agreement is good. On the other hand the spread of experimental
points is still substantial. Polarised quark distributions for up
and down valence quarks as well as non strange, strange quarks and
gluons for $Q^2$ = 1 $\mbox{GeV}^2$ are presented in figure 4.
Dashed curves represent the $+$ and $-$ components for different
parton densities. the solid curves corespond to the difference of
$+$ and $-$ components, the sums of components (not shown)
correspond to nonpolarised parton distributions. We see that
especially polarised gluon distribution function is really tiny
and does not resemble the distribution function for unpolarised
case.

 This function is also quite different from the gluon distribution
(given in \cite{gerst}) used to estimate $\Delta G/G$ in COMPASS
experiment planned at CERN \cite{nassalski}. For $x=0.1$ at $
Q^{2}=1\, {\rm GeV^{2}}$ $\Delta G/G=0.01$ and is below a planned
experimental error. In fig.5 we present spin densities for $u$
quark. We show  $\Delta u_{v}(x,Q^2)$ obtained in the basic fit
(solid line) as well as the same quantity from the fit with
$SU(3)$ symmetric sea (dashed curve) and from the fit where
evolution in $Q^2$ was not taken into account (dotted line). In
the second plot the same curves for $\Delta u(x,Q^2)$ are
presented and one sees from that the values of $\Delta u$ from the
basic fit and fit with $SU(3)$ symmetric sea are very close in
spite of the differences in the valence values. The similar
conclusions are also true for the $d$ quark densities.

\begin{figure}[h]
\noindent \hspace{-0.3cm} \epsfxsize =395pt \epsfbox{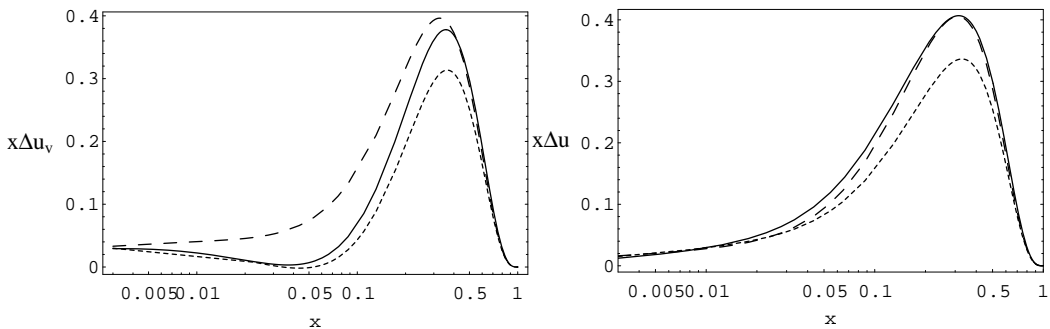}

\caption{\em{Our predictions for $x\Delta u_v(x, Q^{2}=1\, {\rm
GeV^{2}})$ and $x\Delta u(x, Q^{2}=1\, {\rm GeV^{2}})$ in
different models. The solid curve is gotten using the parameters
of our basic fit whereas dashed and dotted lines correspond to the
quantities calculated in the fit with a SU(3) symmetric sea and a
fit without QCD evolution, respectively.}}
\end{figure}

Fixing the value of $a_8$ is very important for the fit. When we
relax the condition for $a_8=0.58$ we get $\chi^2=340.8$, so this
number practically does not change. We get the fit with the
parameters not very different from our basic fit but with very
small $a_8 =0.03$ and positive $\Delta s=0.01$. It seems that
$\Delta s$ is not well determined from the data on spin
asymmetries alone but that does not influences strongly the values
of $\Delta u$, $\Delta d$ and $\Delta G$.

In order to check how the fit depends on the assumptions made
about the sea contribution we have also made fit with $\Delta
\bar{u}=\Delta \bar{d}=2\Delta \bar{s}$, the assumption that
follows directly from MRST unpolarised fit. The $\chi^2$ value
increases significantly and per degree of freedom one gets a
number $\chi^{2}/N_{DF}$ =$\frac{353.2}{418-9}$ =0.86 which is
worse than in our basic fit. In this case we have $\Delta u =
0.80$ ($\Delta u_v = 0.87$, $2\Delta \bar{u} = -0.07$), $\Delta d
= -0.61$ ($\Delta d_v = -0.40$, $2\Delta \bar{d} = -0.21$),
$\Delta s = - 0.07$, $a_0 = 0.11$,  $\Delta G = 0.07$ and
$a_{8}=0.33$. The quantity $\Delta s$ must be negative in order to
get experimental value for $a_8$ and because of our assumption
$\Delta \bar{u}=\Delta \bar{d}=2\Delta \bar{s}$ (the value of
$a_{8}$ does not come out correctly in the fit because of our
assumption $ c_{1s}=c_{1}$) we obtain negative values of non
strange sea for up and down quarks. One sees that the values for
sea polarisation depend very strongly on assumptions we made(in
many papers \cite{grv,alt0,gerst} SU(3) symmetric sea is assumed
that also together with fixing of $a_8$ value gives negative non
strange sea). On the other hand $\Delta u=\Delta u_v+2\Delta
\bar{u}$ and $\Delta d=\Delta d_v+2\Delta \bar{d}$ practically do
not change (however, $\Delta u_v$ and $\Delta d_v$ also change).
Also $\Delta G$ does not change and is small. We get that the
value of $a_{8}$ is not coming out correctly from the fit to spin
asymmetries. Fixing $a_{8}$ and making specific assumption about
$\Delta s$ introduces shifts in nonstrange sea polarisation (and
so in $\Delta u_v$ and $\Delta d_v$) but $\Delta u$, $\Delta d$ do
not change. Because the value of $\Delta s$ is needed to determine
$a_0$ we decided to use additional free parameters for strange sea
contribution in order to determine it (with fixed value of
$a_{8}$) from the fit to experimental data.

 In many papers making fits to experimental data on spin
 asymmetries the assumption of $SU(3)$ symmetric sea was made. We
 have also for comparison made fit with this assumption. The value
 of $\chi ^{2}=350.6$  is substantially higher in comparison
 with our basic fit. In this case we have $\Delta u = 0.81$ ($\Delta
u_v = 0.87$, $2\Delta \bar{u} = -0.06$), $\Delta d = -0.57$
($\Delta d_v = -0.40$, $2\Delta \bar{d} = -0.17$), $\Delta s = -
0.11$, $a_0 = 0.13$,  $\Delta G = 0.14$, $a_{3}=1.38$ and
$a_{8}=0.47$. In this case we also have shifts in values of
valence and sea contributions similar to the case discussed above.

Looking at the dependence of unpolarised quark and gluon densities
we see that after elimination of most singular term in  $\Delta
d_v(x)$  the most singular behaviour for small $x$ one has for
$\Delta u_v(x)$ and $\Delta M(x)$. For comparison we have
investigated the model when in polarised densities these singular
contributions are absent. In this case $\Delta u_v$ and $\Delta M$
are $\sqrt{x}$ less singular than in our basic fit. For such a fit
we get $\chi^{2}/N_{DF}$ =$\frac{356.6}{418-8}$ =0.87, i.e
significantly higher than in our basic fit. We get in this case:
$\Delta u = 0.77$ ($\Delta u_v = 0.57$, $2\Delta \bar{u} = 0.20$),
$\Delta d = -0.38$ ($\Delta d_v = -0.63$, $2\Delta \bar{d} =
0.25$), $\Delta s = - 0.10$, $a_0= 0.28$, $\Delta G = 0.22$. In
such fit the integrated quantities taken over the whole range of
$0 \leq x \leq 1$ and in the truncated one ($0.003 \leq x \leq 1$)
differ very little. The quantity $\Delta G$ is positive and
different from zero. So it is possible to get the fit with
practically no change of integrated quantities in the region
between $x=0$ and $x=0.003$ but with significantly higher
$\chi^{2}$ value. For $Q^2$=1 $\mbox{GeV}^2$ we have $\Gamma_1^p =
0.122$ and $\Gamma_1^n = -0.041$.

The obtained results can be compared with the fit when instead of
417 points for different $x$ and $Q^2$ values we take spin
asymmetries for only 160 data points with the averaged $Q^2$
values for the same $x$ (one has smaller errors in this case). In
such fit the ratio of number of neutron to number of deuteron and
proton data points is increased. It seems that the influence of
neutron points is stronger than in basic fit $\chi^{2}/N_{DF}$
=$\frac{118.3}{161-10}$ = 0.78 is a little bit better than in our
basic fit. The integrated values for quark
 and gluon densities are:
$\Delta u = 0.79$ ($\Delta u_v = 0.65$, $2\Delta \bar{u} = 0.14$),
$\Delta d = -0.66$ ($\Delta d_v = -0.60$, $2\Delta \bar{d} =
-0.06$), $\Delta s = - 0.22$, $a_0 =-0.09$, $\Delta G = 0.31$ and
$a_3=1.45$. We see that averaging over $Q^2$ and different numbers
of data points leads to  very similar fit. The values for
integrated valence densities and nonstrage sea contribution are
only a  bit shifted ($\Delta u$ and $\Delta d$ in the whole region
of integration do not differ from the basic fit for non averaged
data and the  same is also true for integrated quantities in the
region $0.003 \leq x \leq 1$). Integrated gluon density is
relatively small and positive. A little bit higher value for
$\Delta G=0.31 \pm 0.28$ we do not consider as significant
difference. $\chi^{2}/N_{DF}$ is very good and smaller then in
\cite{leader} where the same experimental data sample  and  MRST
parton distributions (modified for small values of $x$) were used.

As was already mentioned before we have also made for comparison
fits neglecting evolution of parton densities with $Q^2$ (formulas
from the simple parton model). We get for non averaged data sample
$\chi^{2}/N_{DF}$ =$\frac{349.9}{418-9}$=0.86 (biger than in our
basic fit: $\chi^{2}/N_{DF}$ =0.84): $\Delta u = 0.66$ ($\Delta
u_v = 0.56$, $2\Delta \bar{u} =0.10$), $\Delta d = -0.49$ ($\Delta
d_v = -0.49$, $2\Delta \bar{d} =0.0$), $\Delta s = - 0.20$, $a_0
=-0.03$, $a_{3}=1.14$, $\Gamma_1^p = 0.108$, $\Gamma_1^n$ =
-0.082.
 For averaged data points we get $\chi^{2}/N_{DF}$
=$\frac{125.4}{161-9}$=0.83 (this number should be compared with
$\chi^{2}/N_{DF}$ =0.78, the corresponding quantity from the NLO
fit) and we have: $\Delta u = 0.66$ ($\Delta u_v = 0.58$, $2\Delta
\bar{u} = 0.08$), $\Delta d = -0.48$ ($\Delta d_v = -0.48$,
$2\Delta \bar{d} =0.0$), $\Delta s = - 0.20$, $a_0 =-0.03$. Hence,
$\chi^2$ per degree of freedom is smaller in the case of averaged
sample. We see that both fits give very similar results. It means
that the averaging of data does not influence the fit when we do
not take $Q^{2}$ evolution into account (the differences are also
very small in the $0.003 \leq x \leq 1$ region).

It has been pointed out \cite{alt0} that the positivity conditions
could be restrictive and influence the contribution of polarised
gluons. We have also made a fit to experimental data without such
assumption for polarised partons. The $\chi ^2$ value does not
changed much $\chi^{2}/N_{DF}$ =$\frac{340.7}{418-10}$ =0.84 and
we get $\Delta u = 0.84$ ($\Delta u_v = 0.72$, $2\Delta \bar{u} =
0.12$), $\Delta d = -0.74$ ($\Delta d_v = -0.50$, $2\Delta \bar{d}
=-0.24$), $\Delta s = - 0.24$, $a_0= -0.13$, $a_{3}=1.57$, $\Delta
G =0.02$. The results are a little bit different but the value of
$\Delta G$ is not influenced  by the positivity conditions. The
same is also true in the case of averaged data. It seems that our
positivity conditions are not very restrictive.

We have made fits for two samples of data with averaged $Q^2$
values and non averged ones (adding neutron data from E154 and
Hermes experiments) leading to very similar results for calculated
parameters (except small difference in $\Delta G$). The value of
$a_3$ was not fixed in the fit and comes out high in comparison
with  experimental value. In order to check the influence of
different assumptions about strange sea we considered fits without
fixing $a_8$ value, with $SU(3)$ symmetric sea and with modified
sea contribution. The models with less singular behaviour for
valence $u$ quark and sea contribution were also discussed. In
most of the modifications the $\chi^{2}$ value increases
significantly. For comparison we have also considered fits to the
simple parton model neglecting $Q^2$ dependence of parton
densities. It seems that splitting of integrated densities $\Delta
u$, $\Delta d$ into valence and sea contribution  is model
dependent ($\Delta u$ and $\Delta d$ do not differ much). The
integrated gluon contribution comes out small. The best fits
(measured by $\chi^2$ per degree of freedom) we have for zero (for
non averaged data points) or rather small (for averaged data)
gluon polarisation. It seems that from results of our fits the
perspective of measuring in COMPASS $\Delta G/G$ is not very
encouraging. The experimental accuracy still must be improved and
probabely additional experiments are needed in order to make more
precise statements about polarised quark and gluon densities.

\end{document}